# On perturbation theory around the atomic limit of strongly correlated electron systems: a new approach based on Wick's theorem

JAN BRINCKMANN

Institut f. Festkörperphysik,
Techn. Hochschule Darmstadt, D-64289 Darmstadt, Germany

**Abstract**

A new perturbational approach to spectral and thermal properties of strongly correlated electron systems is presented: The Anderson model is reexamined for $U \to \infty$, and it is shown that an expansion of Green's functions with respect to the hybridization $V$ built on Feynman diagrams obeying standard rules is possible. The local correlations of the unperturbed system (the atomic limit) are included exactly through a two-particle vertex. No auxiliary particles are introduced into the theory. As an example and test the small energy scale and many-body resonance of the Kondo problem are reproduced analytically.

**Correspondence to:**    J. Brinckmann
Inst. für Festkörperphysik,
Technische Hochschule Darmstadt
Hochschulstr. 6,
D-64289 Darmstadt
Fed. Rep. of Germany

Fax.: ++49-(0)6151 / 163681

e-mail: jan@spy.fkp.physik.th-darmstadt.de

For the present a conventional perturbation theory around the atomic limit of Anderson or Hubbard models, based on Feynman diagrams is defeated by the ionic interaction Hamiltonian that has to be dealt with. In 'slave boson' techniques Wick's theorem is recovered via auxiliary particles, but for the price of a constraint [1, 2] to be maintained. It leads, if taken strictly [3], to the unconventional time ordered perturbation theory [4, 5, 6]. Other approaches [7] start from commutation relations of Hubbard operators [8, 5] and aim at an adequate 'Wick's theorem', which appears as a recurrance relation and does not provide with strict diagram rules [9]. It will be demonstrated in the following for the $f$-Green's function of the $U \to \infty$ impurity Anderson model [10] that invariance properties in certain Fock-space sectors allow for an expansion in the hybridization $V$ using Wick's theorem, while strong local correlations are included exactly.

Introducing Hubbard operators, the model with $N$ angular momentum states $m$ reads:

$$H = \sum_{k,m} \varepsilon_{km} c^\dagger_{km} c_{km} + \sum_m \varepsilon^f_m X^{mm} +$$
$$+ \sum_{k,m} \left( V_k X^{m0} c_{km} + V_k^* c^\dagger_{km} X^{0m} \right) .$$

The $X$-operators act on Fock-space states $|m_1, \ldots, m_n; N_c\rangle$ charcterized by $0 \leq n \leq N$ occupied local $f$-states and some set $N_c$ of conduction band occupation numbers. They cause only a few non-vanishing transitions in the subspace with $n \leq 1$:

$$X^{mm'}|m_1; N_c\rangle = |m; N_c\rangle \delta_{m',m_1} ,$$
$$X^{0m}|m_1; N_c\rangle = |(0); N_c\rangle \delta_{m,m_1} , \quad (1)$$
$$X^{m0}|(0); N_c\rangle = |m; N_c\rangle .$$

A thermal expectation value, e.g. the internal energy $I = \text{tr}'[e^{-\beta H} H]/Z$ includes only this restricted Hilbert space in its trace (denoted by the prime), therefore it is sufficient to consider matrixelements like $\langle M'; N'_c|e^{-\beta H} H|M; N_c\rangle$ with $M, M' = 0, m$. The $X$-operators influence subsequently created intermediate states according to the relations (1), but actually those are conserved while replacing $X^{mm'}$ by canonical $f$-electrons $f^\dagger_m f_{m'}$ and $X^{0m}$ by $f_m$, as far as $X^{m0}$ is not altered. The term in the primed trace may then be simplyfied according to $I = \text{tr}'[e^{-\beta \widetilde{H}} \widetilde{H}]/Z$ with

$$\widetilde{H} = \widetilde{H}^0 + \sum_{k,m} \left( V_k X^{m0} c_{km} + V_k^* c^\dagger_{km} f_m \right) ,$$

$$\widetilde{H}^0 = \sum_{k,m} \varepsilon_{km} c^\dagger_{km} c_{km} + \sum_m \varepsilon^f_m f^\dagger_m f_m .$$

In general this feature holds for any operator $\mathcal{O}$

$$\langle M'; N'_c|\mathcal{O}(\{X^{m0}, X^{0m}, X^{mm'}\})|M; N_c\rangle =$$
$$= \langle M'; N'_c|\mathcal{O}(\{X^{m0}, f_m, f^\dagger_m f_{m'}\})|M; N_c\rangle$$

in the subspace with total ionic occupation number $n = 0$ or $n = 1$. Note that matrixelements of a hermiteian operator like $O = (e^{-\beta H} H)$ remain hermiteian.

The local $f$-propagator is written this way with (imaginary time) Heisenberg operators involving $\widetilde{H}$,

$$F_{m,m'}(\tau, \tau') = -<\mathcal{T}\{X^{0m}(\tau) X^{m'0}(\tau')\}>$$
$$= -\text{tr}' \left[ e^{-\beta \widetilde{H}} \mathcal{T}\{f_m(\tau) X^{m'0}(\tau')\} \right]/Z .$$

It still requires a reduced trace, the quadratic unperturbed Hamiltonian $\widetilde{H}^0$ itself is not sufficient to get Feynman rules instead of (time ordered) Goldstone diagrams [11]. But due to the very existence of $X^{m'0}(\tau')$ we may enlarge the trace to the full Fock space, and inserting $X^{m0} = f_m^\dagger \prod_{m_1 \neq m}(1 - n_{m_1}^f)$, $n_{m_1}^f = f_{m_1}^\dagger f_{m_1}$, we are left with a conventional interacting $N$-particle propagator. For the special case $N = 2$, $m \equiv \sigma = \pm 1$ it reads

$$F_{\sigma,\sigma'}(\tau, \tau') = -\widetilde{Z}/Z \cdot$$
$$\cdot <\mathcal{T}\{f_\sigma(\tau) f_{-\sigma'}(\tau' + 0_+) f_{-\sigma'}^\dagger(\tau') f_{\sigma'}^\dagger(\tau')\}>^{\widetilde{H}}, \quad (2)$$

$\widetilde{Z} = \text{tr}[e^{-\beta \widetilde{H}}]$. Now Wick's theorem is applicable in (2), and the expansion follows a standard treatment (see, e.g. [12]): For $N = 2$, a diagram of order $V^{2n}$ consists of $n$ simple hybridization vertices (Fig. 1 right) and $n$ two-particle vertices (Fig. 1 left). These are linked by bare conduction electrons $G_{k,\sigma}^0(i\omega_l) = 1/[i\omega_l - \varepsilon_{k\sigma}]$, (full lines) and bare 'unprojected' $f$-propagators
$\widetilde{G}_\sigma^0(\tau, \tau') = -<\mathcal{T}\{f_\sigma(\tau) f_\sigma^\dagger(\tau')\}>^{\widetilde{H}_0} \leftrightarrow$
$\widetilde{G}_\sigma^0(i\omega_l) = 1/[i\omega_l - \varepsilon_\sigma^f]$ (dashed lines). Its sign $(-1)^{n_c}$ is given by the number of fermion loops $n_c$, and each internal frequency $i\omega_n$ is accompanied by an $\exp(-i\omega_n 0_+)$ (the exponent differs in sign from standard rules !). The prefactor in the $f$-Green's function Fig. 2, resulting from (2) through

$$<X^{00} + X^{\sigma\sigma}> = <1 - n_{-\sigma}^f> =$$

$$= \frac{1}{Z}\text{tr}'[e^{-\beta \widetilde{H}}(1 - n_{-\sigma}^f)] = \frac{\widetilde{Z}}{Z} <1 - n_{-\sigma}^f>^{\widetilde{H}},$$

already reveals the exact spectral weight belonging to the sum rule

$$\int_{-\infty}^\infty d\omega\, \rho_\sigma^f(\omega) = <\{X^{0\sigma}, X^{\sigma 0}\}>,$$

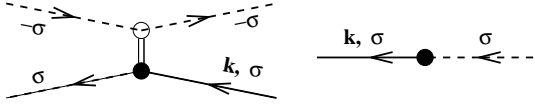

Figure 1: *Vertices resulting from the application of Wick's theorem. Spin, frequency and direction of arrows are conserved. Filled circles represent matrix elements $V_k^{(*)}$. The two-particle vertex (left) permits an f-creation only if a hole of opposite spin is present.*

$\rho_\sigma^f(\omega) = -\frac{1}{\pi}\text{Im}F_{\sigma,\sigma}(\omega + i0_+)$. The two fermion

$$F_{\sigma,\sigma}(i\omega_l) = \frac{<X^{00} + X^{\sigma\sigma}>}{<1 - n_{-\sigma}^f>^{\widetilde{H}}} \cdot \left\{ \begin{array}{c} \text{diagram} \end{array} \right\}$$

Figure 2: *Skeleton representation of the physical local Green's function. Conservation of the incoming frequency is maintained by the dummy vertex $\square$.*

loops emerging in Fig. 2 from contracting three of the four external time labels of (2), contain the overall sign and cancel the prefactor's denominator; in a trivial way in the one-particle part (Fig. 2, first term) leaving only a single fully renormalized 'unprojected' Green's function $\widetilde{G}$ (dashed double line) times the spectral weight factor. The spin restriction at vertices Fig. 1 (left) neither permits internal spin sums nor exchange parts [12] to appear.

As a test case for the perturbation scheme formulated above, 'to see how it works', some important results of the Kondo problem are now rederived by analytical calculation: As might be expected, a mean-field (MF) theory based on a neglect of the irreducible vertex function [12] $\Gamma_C^2$ (Fig. 2, second term) and on a self-consistent treatment of $\widetilde{G}(i\omega_l)$ keeping only its lowest order (spin degenerate) skeleton self energy (see

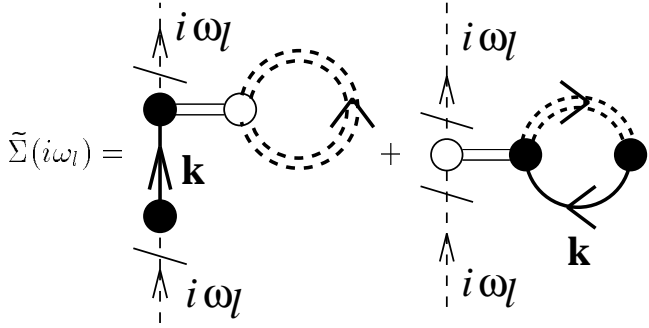

Figure 3: *Lowest order skeletons to the self energy of the 'unprojected' propagator $\widetilde{G}(i\omega_l)$.*

Fig. 3), leads at small temperatur $T$ in the extreme Kondo limit $V \ll D \ll |\varepsilon^f|$ only to a slight shift of the bare $f$-level $\varepsilon^f \to \varepsilon^f + J/2$, $J \equiv V^2/|\varepsilon^f|$ ($\varepsilon^f < 0$, $D$ is half the conduction bandwidth). Actually, the resulting MF-equations are in formal equivalence to those of the slave boson MF problem (see, e.g. [13]) at $N \to 1$. Instead, in ignoring the MF shift, an analytical solution of the lowest order Bethe-Salpeter equations for the particle–particle (pp) channel and for the particle–hole (ph) channel contributing to $\Gamma_C^2$ reproduces the Abrikosov-Suhl resonance [14] in the $f$-spectral function

$$\rho^f(\omega)\Big|_{\substack{T \to 0 \\ |\omega| \ll D}} = <X^{00} + X^{\sigma\sigma}> \frac{\pi T_A}{\Delta}\delta(\omega - T_A),$$

including the proper energy scale [14, 15] $T_A = D \exp\left(\frac{\pi \varepsilon^f}{2\Delta}\right)$. At small energies $\rho^f$ is most influenced by the (pp)-channel which introduces the spin-flip scattering, the constant background due to the (ph)-channel has been left out here. Our result compares very well with analytical treatments in direct perturbation theory based on non-self-consistent resummation [16, 17]: By setting $<X^{00} + X^{\sigma\sigma}> \approx \frac{1}{N} = \frac{1}{2}$ the spectrum of [18] is exactly reproduced.

The discussion above reveals clearly that it is not too difficult to set up an expansion of Green's functions in $V$ around the atomic limit by Feynman diagrams based on Wick's theorem, with no need of 'slave' particles. Also, as a first test, it has been shown that simple resummations can be treated analytically and lead to correct results in the Kondo limit of the Anderson model. Among other questions to be tackled in the future, like improved self-consistent approximations, the study of lattice models could be an interesting field of application for the new approach [19].

**Acknowledgements:**

The author is very grateful to Prof. N. Grewe for many intensive discussions on the subject of this work.